\documentclass[sort&compress,numbers,times,5p,twocolumn,fleqn,letter]{elsarticle}

\journal{a journal for peer review}

\usepackage{times,pslatex}
\usepackage{amsfonts}
\usepackage{amssymb}
\usepackage{amsmath}
\usepackage{graphicx}
\usepackage[all]{xy}
\usepackage{amsthm}
\usepackage{color}
\usepackage{xcolor}
\usepackage{bm}
\usepackage{csquotes}
\usepackage[section]{algorithm}
\usepackage{appendix}

  \RequirePackage{enumerate} 
          \usepackage[shortlabels]{enumitem}
\usepackage[colorlinks, linkcolor=reference,
 citecolor=citation,
         urlcolor=e-mail]{hyperref}
\usepackage[numbers]{natbib}

\RequirePackage{verbatim}

\RequirePackage{hyperref}

\RequirePackage{color}

\definecolor{conditional}{rgb}{0,1,0}
\definecolor{e-mail}{rgb}{0,.40,.80}
\definecolor{reference}{rgb}{.20,.60,.22}
\definecolor{mrnumber}{rgb}{.80,.40,0}
\definecolor{citation}{rgb}{.20,.60,.22}

\usepackage{bbm}

\DeclareMathOperator{\IXa}{[IXa]}
\DeclareMathOperator{\VIIIa}{[VIIIa]}
\DeclareMathOperator{\Xa}{[Xa]}
\DeclareMathOperator{\Va}{[Va]}
\DeclareMathOperator{\APC}{[APC]}
\DeclareMathOperator{\IIa}{[IIa]}

\usepackage{tikz}
\usetikzlibrary{arrows,decorations.markings}

\newtheorem{theorem}{Theorem}

\newtheorem{example}{Example}

\theoremstyle{definition}
\newtheorem{definition}{Definition}
\newtheorem{remark}{Remark}

\def\BibTeX{{\rm B\kern-.05em{\sc i\kern-.025em b}\kern-.08em
    - T\kern-.1667em\lower.7ex\hbox{E}\kern-.125emX}}

\DeclareMathOperator{\ord}{ord}

\begin{document}

\begin{frontmatter}
\author[AO]{Alexey Ovchinnikov}\ead{aovchinnikov@qc.cuny.edu}
\author[AP]{Anand Pillay}\ead{apillay@nd.edu}
\author[GP]{Gleb Pogudin}\ead{gleb.pogudin@polytechnique.edu}
\author[TS]{Thomas Scanlon}\ead{scanlon@math.berkeley.edu}
\address[AO]{CUNY Queens College, Department of Mathematics,
65-30 Kissena Blvd, Queens, NY 11367, USA \\ 
CUNY Graduate Center, Mathematics and Computer Science, 365 Fifth Avenue,
New York, NY 10016, USA}
\address[AP]{University of Notre Dame, Department of Mathematics,  Notre Dame, IN 46556, USA}
\address[GP]{LIX, CNRS, \'Ecole Polytechnique, Institute Polytechnique de Paris, 1 rue Honor\'e d'Estienne d'Orves, 91120, Palaiseau, France}
\address[TS]{University of California, Berkeley, Mathematics Department, Evans Hall, Berkeley, CA, 94720-3840} 
\title{Identifiable specializations for ODE models}

\begin{abstract}
The parameter identifiability problem for a dynamical system is to determine whether the parameters of the system can be found from data for the outputs of the system. 
Verifying whether the parameters are identifiable is a necessary first step before a meaningful parameter estimation can take place. 
Non-identifiability occurs in practical models. 
To reparametrize a model to achieve identifiability is a challenge. 
The existing 
approaches have been shown to be useful for many important examples. 
However, these approaches are either limited to linear models and scaling parametrizations or are not guaranteed to find a reparametrization even if it exists. In the present paper, we prove that there always exists a locally identifiable model with the same input-output behaviour as the original one obtained from a given one by a partial specialization of the parameters.  Our result applies to parametric rational ODE models with or without input, and our algorithm can find non-scaling reparametrizations. As an extra feature of our approach, 
the resulting (at least) locally identifiable reparameterization has the same shape: the monomials in the new state variables in the new model are formed in the same way as in the original model.
Furthermore, we give a sufficient observability condition for the existence of a state space transformation from the original model to the new one.
Our proof is constructive and can be translated to an algorithm, which we illustrate by several examples, with and without inputs.
\end{abstract}
\end{frontmatter}

\section{Introduction}
\subsection{Motivation}
Scientists and engineers often model a process under investigation using a parametric ODE
\begin{equation}\label{eq:mainODE}
\Sigma(\bar\alpha) :=\begin{cases}
\bar x'(t) = \bar f(\bar x(t),\bar\alpha,\bar u(t))\\
\bar y(t) = \bar g(\bar x(t), \bar\alpha, \bar u(t)),
\end{cases}
\end{equation}
which is a system of ordinary differential equations in $\bar x(t)$ involving some unspecified parametric constants $\bar\alpha$.  The unknown parameters are usually determined (identified) from the input $\bar u(t)$ and measured output data $\bar y(t)$. However, for some parametric ODEs, due to their intrinsic structure, it might not be possible to identify the parameters uniquely from the input and measured data (even noise-free). Therefore, while designing a model, it is crucial to make sure that all parameters in a parametric ODE model  are 
identifiable.  In the literature, this type of identifiability is called generic structural identifiability. Since we will only mean ``generic structural'', we will omit saying ``generic structural''.

 There is a more refined distinction of types of identifiability, useful for further analysis of the model and parameter estimation. If only finitely many parameter values fit the data, this parameter is said to be locally identifiable. In case of unique value, it is said to be globally identifiable.

If the initially designed model has non-identifiable parameters, the next natural step would be to find another model with the same input-output behavior but with all parameters identifiable.
This is a problem we study in the present paper.

\subsection{Input-output identifiability vs. identifiability}
The results we show in this paper are for a closely related property called  global/local input-output identifiability. In this point of view, the parameters are determined from input-output equations (the equations relating the inputs and the outputs obtained by eliminating the state variables~\cite{O1990,O1991,DJNP2001,MRCW21}). Global/local input-output identifiability and global/local identifiability are not equivalent, but there are sufficient conditions for the equivalence, see \cite{ident-compare}, which can be checked algorithmically and often  hold in practice~(e.g., \cite[Table~3]{dong2023differential}). 

The standard definitions of input-output identifiability (such as the one we use in the paper) operate under the assumption that the inputs are sufficiently exciting~\cite{Glad1990} meaning that they are assumed not to satisfy any particular differentially-algebraic equations.

\subsection{Prior work}

There exist efficient algorithms for finding reparametrizations of  a specific form such as scaling transformations~\cite{Hubert2013} or linear reparametrizations~\cite{Ovchinnikov2021,JimnezPastor2022}.
More refined results have been obtained for scaling reparametrizations of linear compartmental models~\cite{MESHKAT201446,Baaijens2015OnTE}.
Several approaches have been proposed for producing locally identifiable reparametrizations~\cite{Gunn1997,https://doi.org/10.1002/rnc.5887,JOUBERT2020108328,EC2000} which succeed in finding nontrivial parametrizations for models from the literature but are not guaranteed to produce a reparametrization if it exists. For example, \cite{https://doi.org/10.1002/rnc.5887} will be successful if the chosen ansatz for the Lie symmetries it uses is correct. The method described in \cite{EC2000} extends the  approach from~\cite{Gunn1997} based on solving PDEs resulting from the nullspace of the Jacobian of the output Lie derivatives. It reduces the PDE solving to solving a system of ODEs making it a better fit to computer algebra systems. However, this method is still not a complete algorithm as of now because of the ODE solving step. 
The existence of  an identifiable reparametrisation was also not completely understood: Sussmann's theorem~\cite{Sussmann1976} guarantees the existence of an identifiable model with the same input-output behaviour at the cost of allowing  the models to be defined on a manifold, in other words,  if we permit the replacement of
 ODEs by differential-algebraic equations.

 The paper \cite{MOS2023} gives another approach to find locally and then possibly globally  input-output identifiable reparametrizations by reducing the problem to polynomial system solving over fields of rational functions, which is the main efficiency bottleneck of such an approach in terms of scalability, though the largest non-linear model it was seen being successful has 16 parameters, 9 state variables, 1 input, and 3 outputs.

 An approach from algebraic geometry is used in \cite{FOS2025} to give a procedure to find globally  input-output identifiable reparametrizations of rational ODE models. However, 
 it only provides a complete algorithm for first-order equations.
 For larger systems, this procedure is not fully algorithmic even though being often successful for concrete examples. More work also needs to be done to make this approach scalable to larger systems in terms of efficiency.

\subsection{Our contribution}\label{sec:1:3}
We prove that it is, in fact, always possible to replace the original ODE system with another one with the same input-output behaviour but all the parameters being locally  input-output identifiable by partially specializing parameters, in particular, without changing the ``shape of the system'' (Theorem~\ref{thm:main}).  The latter means that
\begin{itemize}
\item the reparametrized system has the same number of equations and state variables as the old system and 
\item the monomials in the new system are obtained from the monomial of the old system by replacing the old state variable with the new state variables.
\end{itemize}
We also give an example showing that this statement is not true for global identifiability (see Section~\ref{sec:notglobal}).
Under an additional observability condition, we also show that there exists a state space transformation between the original model and the new one (Theorem~\ref{thm:main}).
Our proofs are constructive and can be directly translated into algorithms, which we showcase by several examples (Section~\ref{sec:examples}).  Even though, in our experiments, scaling reparametrizations were frequently sufficient  to achieve (local) IO identifiability, there are cases in which these do not exist. Our algorithm is capable of finding non-scaling reparametrizations as well - see Section~IV.\ref{sec:CRN} for such a non-linear example from chemical reaction networks.
Our {\sc Maple} code for these examples can be found in \cite{exaples-github}.
 We discuss the efficiency of our algorithm in Section~III.\ref{sec:efficiency}.

Finding globally identifiable reparametrizations is important, for instance, to reduce inaccuracy of parameter estimation in optimization-based approaches (see, e.g.,~\cite[tables]{par_est_robust}). Our algorithm guarantees to find a locally identifiable reparametrization. If the reparamentrization it found is locally but not globally identifiable, then, as the next step, one can apply~\cite{MOS2023} or~\cite{FOS2025} to reparametrize further.

  In the approaches bases on input-output equations,
   no consideration is given to the model initial conditions. There is an approach under development via simplifying Lie derivatives to find a   reparametrizations in which all parameters and initial conditions are globally identifiable~\cite{DPR2024}. However, reparametrized models in this approach are in general differential-algebraic systems. Our algorithm cannot tackle identifiability of the initial conditions because  it is based on input-output equations. However, reparametrized models in our approach are still of ODE form~\eqref{eq:mainODE}.

\section{Main result}
\subsection{Preliminaries and setup}\label{sec:notation}
In what follows,  for a letter $z$, $\bar z$ means that $\bar z = (z_1, \ldots, z_n)$ is a tuple of length $n$ (different tuples can have different lengths); we will write expressions like $\bar z \in A$ to mean that 
$z_i \in A$ for each component $z_i$ of $\bar z$.
Our main object will be
an ODE system
\begin{equation}\label{eq:ODEsystem}
\Sigma(\bar\alpha) :=\begin{cases}
\bar x'(t) = \bar f(\bar x(t),\bar\alpha,\bar u(t))\\
\bar y(t) = \bar g(\bar x(t), \bar\alpha, \bar u(t)),
\end{cases}
\end{equation}
where $\bar{\alpha}$ is a vector of scalar parameters, $\bar{x}(t)$, $\bar{y}(t)$, and $\bar{u}(t)$ are the state, output, and input functions, respectively (in what follows we will omit the dependence on $t$ for brevity).
We will focus on rational ODE models, that is, the case of $\bar f$ and $\bar g$ being tuples of rational functions over $\mathbb{Q}$ (or $\mathbb{C}$).

To formally define the main property of interest, input-output identifiability (IO-identifiability), we will introduce some notation from algebra.
\begin{enumerate}
  \item A {\em differential ring} $(R,{}^\prime)$ is a commutative ring with a derivation $':R\to R$, that is, a map such that, for all $a,b\in R$, $(a+b)' = a' + b'$ and $(ab)' = a' b + a b'$. 
  \item The {\em ring of differential polynomials} in the variables $x_1,\ldots,x_n$ over a field $K$ is the ring \[K\left[x_j^{(i)}\mid i\geqslant 0,\, 1\leqslant j\leqslant n\right]\] with a derivation defined on the ring by $\left(x_j^{(i)}\right)' := x_j^{(i + 1)}$. 
  This differential ring is denoted by $K\{x_1,\ldots,x_n\}$.
    \item  For a differential polynomial $p \in K\{x_1,\ldots,x_n\}$, its {\em order} in a variable $x_i$ is the order of the highest derivative of $x_i$ that appears in $p$. We denote this by $\ord_{x_i} p$. The order of $p$ in a tuple of variables $\bar{x}$ is the maximal order of all $x_i$ from $\bar x$ that appear in $p$. We denote this by $\ord_{\bar x}p$.  When speaking about orders of equations, we will also mean orders of the corresponding differential polynomials.
  \item An ideal $I$ of a differential ring $(R,{}^\prime)$ is called a {\em differential ideal} if, for all $a \in I$, $a'\in I$. For $F\subset R$, the smallest differential ideal containing the set $F$ is denoted by $[F]$.
    \item 
  For an ideal $I$ and element $a$ in a ring $R$, we denote $I \colon a^\infty = \{r \in R \mid \exists \ell\colon a^\ell r \in I\}$.
  This set is also an ideal in~$R$.

    \item Given $\Sigma(\bar\alpha)$ as in~\eqref{eq:ODEsystem}, we define the differential ideal of $\Sigma(\bar\alpha)$ as 
    \[
    I_{\Sigma(\bar\alpha)} := [Q\bar{x}'- Q\bar{f},Q\bar{y}-Q\bar{g}]:Q^\infty \subset \mathbb{Q}(\bar{\alpha})\{\bar{x},\bar{y},\bar{u}\},
    \]
    where $Q$ is the common denominator of $\bar f$ and $\bar g$.
    The relations between the inputs and outputs of the system can be found by intersecting this ideal with the corresponding subring:
    \begin{equation}\label{eq:ioeq}
      I_{\Sigma(\bar\alpha)} \cap \mathbb{Q}(\bar{\alpha})\{\bar{y}, \bar{u}\}.
    \end{equation}
    \item For a field $K$, let $\overline K$ denote the algebraic closure of $K$.
\end{enumerate}

Roughly speaking, input-output identifiability is a property for a parameter to be determined from inputs and outputs using IO-equations.
Here is a precise formal definition:

\begin{definition}[IO-identifiability]\label{def:IO-ident}
\begin{enumerate}
\item 
The smallest field $k$ such that $\mathbb{Q} \subset k \subset \mathbb{Q}(\bar{\alpha})$ and  $I_{\Sigma(\bar\alpha)} \cap \mathbb{Q}(\bar{\alpha})\{\bar{y},\bar{u}\}$ is generated (as an ideal or, equivalently, as a differential ideal) by $I_{\Sigma(\bar\alpha)} \cap k\{\bar{y},\bar{u}\}$ is called \emph{the field of IO-identifiable functions}.

\item 
We call $h \in \mathbb{Q}(\bar{\alpha})$  \emph{IO-identifiable} if $h \in k$. We also call $h \in \mathbb{Q}(\bar{\alpha})$  \emph{locally IO-identifiable} if $h$ is in the algebraic closure of the field $k$.  

\item 
The \emph{IO-equations} are defined as the monic characteristic presentation of the differential ideal $I_{\Sigma(\bar\alpha)} \cap \mathbb{Q}(\bar{\alpha})\{\bar{y},\bar{u}\}$ (see \cite[Definition~6 and Section~5.2]{ident-compare} for more details).  For a fixed differential ranking, such a monic characteristic presentation is unique~\cite[Theorem~3]{Boulier2000}.
\end{enumerate}
\end{definition}

In many cases, IO-identifiability is equivalent to identifiability. 
See, e.g., a rigorously written definition of identifiability \cite[Definition~2.5]{HOPY2020}, \cite[Section~4]{ident-compare} for a sufficient condition for the equivalence, and \cite[Examples~2.6 and 2.7]{ident-compare} for simple examples of non-equivalence. 
Additionally, it turns out that IO-identifiability is equivalent to multi-experimental identifiability \cite[Theorem~19]{allident}. 
Finally, several software packages check IO-identifiability~\cite{DAISY,DAISY_IFAC,DAISY_MED,COMBOS,COMBOS2} and find all IO-identifiable functions of  the parameters~\cite{ilmer_web-based_2021}.

\begin{remark}
Some authors prefer to work only with differential fields
$(K,{}')$ containing the field of complex numbers 
$\mathbb{C}$ as a subfield of the field of constants, 
that is, the field of elements $a$ of $K$ satisfying 
$a'=0$.  With this convention, the field of 
IO-identifiable functions is taken to be the smallest
field $k$ for which $\mathbb{C} \subset k \subset \mathbb{C}(\bar{\alpha})$.  For computational reasons, 
we prefer to work over $\mathbb{Q}$ instead of 
$\mathbb{C}$. All of what we discuss in this paper
may be generalized to the case of $\mathbb{C}$ as the 
base with no essential changes to the arguments.
\end{remark}

\subsection{Statement of the main result}
\begin{theorem}\label{thm:main}
 For every ODE system $\Sigma(\bar{\alpha})$ as in~\eqref{eq:ODEsystem}, there exists a tuple $\widetilde\alpha \in \overline{\mathbb{Q}(\bar{\alpha})}$
of the same length as $\bar{\alpha}$ such that
\begin{itemize}

\item the entries of $\widetilde{\alpha}$ are 
are locally IO-identifiable;
\item the system $\Sigma(\widetilde{\alpha})$ has the same input-output equations as the original $\Sigma(\bar{\alpha})$.
\end{itemize}
Furthermore,  let $E_1,\ldots, E_m$ be the IO-equations of $\Sigma(\bar\alpha)$. If ${\sum_i \ord_{\bar y} E_i}$ is equal to the dimension of the model ($n$ in the notation of~\eqref{eq:ODEsystem}),
 then the state variables of $\Sigma(\widetilde\alpha)$ can be expressed as algebraic functions of $\bar x$ and $\bar\alpha$.

Thus, system $\Sigma(\widetilde{\alpha})$ is a locally identifiable reparametrization of system $\Sigma(\bar\alpha)$, and these systems have the same shape (see Section~\ref{sec:1:3}).
\end{theorem}

\begin{remark} If the ODE system comes with additional constraints on the parameters $\bar\alpha$ (e.g., all parameters are non-negative), then one can use standard interval arithmetic algorithms and their implementations to find the corresponding constraints for $\widetilde\alpha$. 
\end{remark}

\subsection{On the existence of the state transformation}

Let us make some remarks about the second part of the theorem.

\begin{remark}[On the last condition and observability]
    The condition in Theorem~\ref{thm:main} that the sum of the orders of the input-output equations  w.r.t. $\bar{y}$ be equal to the dimension of the system is, in fact, equivalent to the fact that all the states are locally observable (the initial conditions are locally identifiable) if the parameters are assumed to be \emph{known} (see \cite[Proposition~5]{G1990}). This equivalence can also be deduced from~\cite[Corollary~4.11]{Hubert2} and~\cite[Proposition~3.4]{HOPY2020}. 
    In particular, this restriction is significantly milder than observability of all the states.
\end{remark}

\begin{example}[Non-existence of the state transformation]
    Let us give an example showing that the condition for the orders of IO-equations summing up to $n$ cannot be removed. 
    Consider the system
    \[
    \begin{cases}
        x_1' = \alpha_1,\\
        x_2' = \frac{x_2}{\alpha_2},\\
        y = x_1.
    \end{cases}
    \]
    The IO-equation of this model is $y' - \alpha_1 = 0$, so only $\alpha_1$ is (locally) IO-identifiable. 
    Then any specialization described in Theorem~\ref{thm:main} will be of the form
    \[
    \begin{cases}
        w_1' = \alpha_1,\\
        w_2' = f(\alpha_1) w_2,\\
        y = x_1.
    \end{cases}
    \]
    for some nonzero algebraic function $f \in \overline{\mathbb{Q}(\alpha_1)}$.
    Any solution of the new system will be of the form \[\left(\alpha_1 t + c_1,\ c_2 e^{f(\alpha_1)t}\right),\] while any solution of the original system was \[\left(\alpha_1 t + c_3,\ c_4 e^{\alpha_2t}\right).\]
    The existence of an algebraic state space transformation as in the theorem would imply that $e^{f(\alpha_1)t}$ is algebraic over $\mathbb{Q}(\alpha_1, e^{\alpha_2 t})$, which is not the case.
\end{example}

\begin{remark}[On possible preprocessing]\label{rem:preprocessing}
    While not all the models satisfy the condition of the second part of the theorem, one can use the approach from~\cite[Section~3]{Forsman} (see~\cite[Section~3.2]{Pavlov2022} for the case with inputs) by constructing a realization of the input-output equation of the model of minimal dimension.
    The corresponding system~\eqref{eq:xw} from our proof will be nonsingular in this case, providing a coordinate change.
    After that, Theorem~\ref{thm:main} can be applied.
    Note that since the dimension of the model changes under this transformation, it is not possible to preserve the shape (see Section~\ref{sec:1:3}) as in Theorem~\ref{thm:main}.
    We give an example of such  a preprocessing in Section~\ref{sec:CRN}.
\end{remark}

\section{Constructive proof of Theorem~\ref{thm:main}  and efficiency}
\label{sec:1}
\subsection{Constructive proof}
We break down the proof into several {\bf steps}, each of which can be viewed as a step in an algorithm to solve the problem:
\begin{enumerate}
\item
We first fix 
$\bar{\beta} = (\beta_1, \ldots, \beta_N) \in \mathbb{Q}(\bar{\alpha})$ 
 a tuple generating
the field of IO-identifiable functions of an ODE system $\Sigma(\bar{\alpha})$
and write 
 it
explicitly in terms of~$\bar{\alpha}$:
\[
  \beta_1 = p_1(\bar{\alpha}),\; \ldots, \;\beta_N = p_N(\bar{\alpha})
\]
for  some rational functions $p_1, \ldots, p_N \in \mathbb{Q}(\bar{\alpha})$.

\item
Let $n$ be the order of the original ODE system, that is, the length of the tuple $\bar{x}$. 
Let $E_{1}(\bar\alpha,\bar u)(\bar y), \ldots, E_m(\bar{\alpha}, \bar{u})(\bar{y})$ be  the input-output equations with respect to any ranking on $\bar{y}$.
Then the orders of $\bar{y}$ and $\bar{u}$ in $E_{i}$ do not exceed $n$ for every $1 \leqslant i \leqslant m$.

Using Lie derivatives, we can write $\bar{y}, \ldots, \bar{y}^{(m)}$ as rational functions $R_0,\ldots,R_m$ in $\bar{x}, \bar{\alpha}, \bar{u}, \ldots, \bar{u}^{(m)}$  for any $m$. 
Therefore, for all $m \geqslant n$, $\bar{y}, \ldots, \bar{y}^{(m)}$ are algebraically dependent over $\mathbb{Q}(\bar{\alpha})\langle \bar{u}\rangle$, where $F\langle a \rangle$ denotes the differential field generated by $a$ over $F$, that is, $F(a, a', a'', \ldots)$.    Furthermore, by~\cite[Lemma~3.18]{HOPY2020}, we have
\begin{equation}\label{eq:fingen}
\mathbb{Q}(\bar{\alpha})\langle \bar{y}, \bar{u} \rangle = \mathbb{Q}\big(\bar{\alpha}, \bar{y}, \bar{y}', \ldots, \bar{y}^{(m)}\big) \langle \bar{u} \rangle.
\end{equation}
Let $M$ be the Jacobian matrix of $ \bar R :=  R_0,\ldots,R_m$ with respect to $\bar{x}$.
Then, by the Jacobian criterion together with~\eqref{eq:fingen}, its rank $r$ will be equal to \[\operatorname{trdeg}_{\mathbb{Q}(\bar{\alpha})\langle \bar{u}\rangle} \mathbb{Q}(\bar\alpha)\langle \bar{u}\rangle (\bar{R}) =\operatorname{trdeg}_{\mathbb{Q}(\bar{\alpha})\langle \bar{u}\rangle} \mathbb{Q}(\bar\alpha)\langle\bar{u}\rangle\langle \bar{y}\rangle,\]
 where $\operatorname{trdeg}$ means the transcendence degree of the field extension.
Let $D$ be the determinant of a nonsingular $r \times r$-minor of $M$.
We consider $D$ as a rational function in $\bar{x}, \bar{u}, \bar{u}', \ldots, \bar{u}^{(n)}$, and take any nonzero coefficient of its numerator, which we denote by $D_0(\bar{\alpha})$. 
 We will use this coefficient in the next step.
\item\label{step:3}
We now form a system of algebraic equations and inequations ($\ne$) in $\widetilde{\alpha}$ over $\mathbb{Q}(\bar{\beta})$:
\begin{equation}\label{eq:alphatilde}
\begin{cases}
    \beta_1 = p_1(\widetilde{\alpha}),\\
    \vdots\\
    \beta_N = p_N(\widetilde{\alpha}),\\
    D_0(\widetilde{\alpha})\cdot C(\widetilde\alpha) \neq 0,
\end{cases}
\end{equation}
where $C$ is the common denominator of all coefficients from $\mathbb{Q}(\bar\alpha)$ in~\eqref{eq:ODEsystem}.
System~\eqref{eq:alphatilde} has a solution $\widetilde{\alpha} = \bar{\alpha}$ and, thus, by Hilbert's Nullstellensatz, has a solution $\widetilde{\alpha}$ in the algebraic closure of the ground field, $\overline{\mathbb{Q}(\bar{\beta})}$.
For this solution, we have $\bar{\beta} \in \mathbb{Q}(\widetilde{\alpha})$ and $\widetilde{\alpha} \in  \overline{\mathbb{Q}(\bar{\beta})}$ by construction.    
Furthermore, since the rank of $M|_{\bar{\alpha} = \widetilde{\alpha}}$ is also $r$, the specialized ODE system has the same input-output equations.

 Indeed, the specialization is possible because ${C}(\widetilde\alpha)\ne 0$.
The specializations of the input-output equations $E_1(\widetilde{\alpha}, \bar{u})(\bar{y}), \ldots, E_m(\widetilde{\alpha}, \bar{u})(\bar{y})$ belong to the specialized elimination ideal $J := I_{\Sigma(\widetilde{\alpha})} \cap k\{\bar{y}, \bar{u}\}$ and they still form a characteristic presentation of some prime differential ideal, we denote it by $J_0$.
 Therefore, if $J_0 = J$, then $E_1(\widetilde{\alpha}, \bar{u})(\bar{y}), \ldots, E_m(\widetilde{\alpha}, \bar{u})(\bar{y})$ are indeed the input-output equations of the specialized system $\Sigma(\widetilde{\alpha})$.
 
 Assume the contrary, that $J_0$ is a proper subset of $J$.
 Since $J \cap k\{\bar{u}\} = J_0\cap k\{\bar{u}\} = \{0\}$ by~\cite[Lemma~3.1]{HOPY2020},
 the sum of the orders of the characteristic presentation of $J_0$ with respect to $\bar{y}$ must be less than $\sum_{i = 1}^m \ord_{\bar{y}} E_i = r$.  This is because, if one prime ideal is contained in another one and they have the same dimension, then the ideals are equal.
 On the other hand, under the specialization to $\widetilde\alpha$, the matrix rank is preserved (as shown above), the 
sum of the orders
of the input-output equations of~\eqref{eq:ODEsystem} after specializing $\bar\alpha \to \widetilde\alpha$ is also $r$.
This provides us a contradiction with the assumption that $J_0 \neq J$.
\item
Finally,  let $\bar w$ denote the state variables of $\Sigma(\widetilde\alpha)$.  Assume that the sum of the orders of IO-equations is equal to $n$. 
We will now see how  $\bar  w$ can be obtained from the original $\bar x$.
Consider the irreducible affine variety $V$ defined by the IO-equations in the space with coordinates $\bar y,\ldots, \bar y^{(m)}$ over the field $\mathbb{Q}(\bar\beta)\langle\bar u\rangle$.
The dimension of this variety is equal to sum of the orders of the IO-equations~\cite[Corollary~4.11]{Hubert2}, and thus it is equal to $n$.
The equalities \[\bar y^{(i)} = R_i(\widetilde\alpha,\bar w,\bar u),\ \ i=0,\ldots,m,\]
 define a dominant rational map $\psi$ from the affine $n$-space with coordinates $\bar w$ to $V$ over the field $\overline{\mathbb{Q}(\bar\beta)}\langle
 \bar u\rangle = \overline{\mathbb{Q}(\widetilde\alpha)}\langle\bar u\rangle$. 
 
 Consider the field  $F := \overline{\mathbb{Q}(\bar\beta)}\langle\bar u\rangle(V) = \overline{\mathbb{Q}(\widetilde\alpha)}\langle\bar u\rangle(V)$ and the generic point $\bar Y=(Y_0,\ldots,Y_m)$  of  the variety $V$  with points whose components belong to the field $F$. Since the morphism $\psi$ is dominant and $\bar Y$ is generic, the system of equations $\psi(\bar w) = \bar Y$  in the variables $\bar w$ has a solution in $\overline{F}$.
On the other hand, the field \[\overline{\mathbb{Q}(\bar\alpha)}\langle\bar u\rangle\left( R_0(\bar\alpha,\bar x,\bar u),\ldots, R_m(\bar\alpha,\bar x,\bar u)\right)\] is isomorphic to the extension of $F$ by $\overline{\mathbb{Q}(\bar\alpha)}$  via
\[
R_i(\bar\alpha,\bar x, \bar u) \mapsto Y_i,\ \ 0\leqslant i\leqslant m
.\] Denote this isomorphism by $\varphi$. 
The desired correspondence between $\bar w$ and $\bar x$ can now be found by  applying $\varphi^{-1}$ to a solution of $\psi(\bar{w}) = \bar{Y}$.
This way, we get algebraic functions $\bar{w} = \bar{w}(\bar{\alpha}, \bar{x}, \bar{u}, \bar{u}', \ldots)$ such that 
\begin{equation}\label{eq:xw}
R_i(\widetilde{\alpha}, \bar{w}, \bar{u}) = R_i(\bar{\alpha}, \bar{x}, \bar{u}) \quad \text{ for }i = 0, \ldots, m.
\end{equation}
We will now show that $w_i'$ is indeed equal to $f_i(\bar{w}, \widetilde{\alpha}, \bar{u})$.
By differentiating~\eqref{eq:xw}, we establish that $w_1', \ldots, w_n'$ satisfy the following linear system (cf. \cite[Section~3]{Forsman} and~\cite[Section~3.2]{Pavlov2022}):
\begin{multline*}
\sum\limits_{j = 1}^n w_j' \frac{\partial R_i(\widetilde{\alpha}, \bar{w}, \bar{u})}{\partial w_j} = \sum\limits_{j = 1}^n f_j(\bar{x}, \bar{\alpha}, \bar{u}) \frac{\partial R_i(\bar{\alpha}, \bar{x}, \bar{u})}{\partial x_j}  \\+ \sum\limits_{j = 1}^\ell u_j' \left(\frac{\partial R_i(\bar{\alpha}, \bar{x}, \bar{u})}{\partial u_j} - \frac{\partial R_i(\widetilde{\alpha}, \bar{w}, \bar{u})}{\partial u_j}\right)
\end{multline*}
 for $i = 0, \ldots, m$.
Since $\dim V = n$, the rank of the matrix of this linear system is also $n$, so it has a unique solution.
On the other hand, $f_1(\bar{w}, \widetilde{\alpha}, \bar{u}), \ldots, f_n(\bar{w}, \widetilde{\alpha}, \bar{u})$ is a solution of this system by construction of the $R_i$'s.
Therefore, \[w_i' = f_i(\bar{w}, \widetilde{\alpha}, \bar{u}),\ \ 1\leqslant i\leqslant n,\] for the found algebraic functions $\bar{w} = \bar{w}(\bar{\alpha}, \bar{x}, \bar{u}, \bar{u}', \ldots)$.

Finally, we will show that $\bar{w}(\bar{\alpha}, \bar{x}, \bar{u}, \bar{u}', \ldots)$ in fact does not depend on $\bar{u}$ (or any of its derivatives).
Let $h$ be the largest order of $\bar{u}$ occurring in these expressions, say $u_1^{(h)}$ occurs in $w_1$.
Then $w_1'$ will depend nontrivially on $u_1^{(h + 1)}$.
On the other hand, $w_1' = f_1(\bar{w}, \tilde{\alpha}, \bar{u})$, where $\widetilde{\alpha}$ does not involve $\bar{u}$, $\bar{u}$ is of order zero in $\bar{u}$, and $\bar{w}(\bar{\alpha}, \bar{x}, \bar{u}, \bar{u}', \ldots)$ is of order at most $h$ in $\bar{u}$.
Thus, we have arrived at a contradiction with our original assumption that $\bar{w}(\bar{\alpha}, \bar{x}, \bar{u}, \bar{u}', \ldots)$ depended in $\bar{u}$.

To carry this out computationally, we solve the system of rational equations~\eqref{eq:xw} for $\bar w$ over $\mathbb{Q}(\bar\alpha{, \bar{x}})\langle\bar u\rangle$.
This can be done, for instance, by computing a Gr\"obner bases of the numerator of~\eqref{eq:xw} with an elimination monomial ordering $w_i > \bar x$ for each $i$, $1\leqslant i \leqslant n$, over $\mathbb{Q}(\bar\alpha)\langle\bar u\rangle$ to find a polynomial equation $P(w_i,\bar x,\bar\alpha,\bar u)=0$ for each $i$ (see Sections~\ref{sec:choices},~\ref{sec:LV}, and~\ref{sec:CRN} for concrete examples).
\end{enumerate}

\begin{remark}\label{rem:sim}  The equations given by $p_1,\ldots, p_N$ in~\eqref{eq:alphatilde} can look 
very complicated. Our {\sc Maple} code \url{https://github.com/pogudingleb/AllIdentifiableFunctions} based on~\cite{allident} can  be used to significantly simplify the system by applying the functions {\tt FieldToIdeal} and then {\tt FilterGenerators} to $p_1(\bar \alpha),\ldots,p_N(\bar\alpha)$.
\end{remark}

\begin{remark}
For some models, IO-equations do not depend on a subset of parameters. These parameters are then definitely non-identifiable. From the measurement point of view, these parameters do not affect the differential equations satisfied by the inputs and outputs of the model but can (and typically will) affect the initial conditions of these equations. In particular, a change in these nonidentifiable parameters can typically be ``compensated'' (in the sense of preserving the input and output data) by a change in one of the initial conditions for the states. For example, scaling $b \to \lambda b$ in~\eqref{eq:LV} can be compensated by additionally scaling $x_2(0) \to \tfrac{x_2(0)}{\lambda}$.
\end{remark}

\begin{remark}
The choice of $\widetilde\alpha$ in Step~\ref{step:3} is constrained by~\eqref{eq:alphatilde}. One can then use the standard symbolic tools based on Gr\"obner bases of simplify and solve~\eqref{eq:alphatilde}. In practice, in the examples we presented, we could solve this system by hand as it was simple even for complicated ODE models. For the purposes of this paper, in our examples, we picked  solutions that look the simplest mathematically. However, in practice, the user might have other perspectives when picking solutions after the algorithmic step of solving has found the entire family of all solutions.
\end{remark}

\subsection{Efficiency discussion}\label{sec:efficiency}
We will now discus the efficiency of the algorithm based on the above constructive proof. The first step requires computing IO equations, which is computationally challenging. However, the efficiency of this has been recently improved, and IO equations are computed relatively quickly for such models as MAPK with 12 state variables, 22 parameters, and 5 outputs \cite{dong2023differential}, which was out of reach before.

Expressing Lie derivatives as rational functions of the states, parameters, and inputs in the second step is computationally challenging due to the potentially high size of the expressions.
Finding a non-singular minor of the resulting large symbolic Jacobian matrix is typically computationally challenging too. However, in 
 the examples we have tried, these two steps have not been efficiency bottlenecks once IO equations have been computed. In future  implementations, one could analyze how lazy evaluations of the Lie derivatives could help speed up these steps.

The simplification mentioned in Remark~\ref{rem:sim} can take some time to complete. However, it has never taken too much time in our experiments (several minutes on a laptop even for the larger of our examples, in {\sc Maple} mentioned above or in the Julia implementation \href{https://github.com/SciML/StructuralIdentifiability.jl}{StruncturalIdentifiability.jl}.).
Moreover, after the simplification, in all of the examples we considered, system~\eqref{eq:alphatilde} has  been relatively simple to find a solution of.

The step of finding the state variable transformation that results in the reparametrized ODE system looks potentially challenging because it involves a Gr\"obner basis computation over a field of rational functions. However, in practice, this step is fast as typically we see  affine 
changes of variables, while more elaborate cases are constructed artificially \cite[Section~V.H]{MOS2023}.

We illustrate in Section~VI.\ref{sec:Akt} using a biological example that the computation of IO equations may be the main bottleneck as the rest of the steps of the algorithm result in much smaller computation even if the IO equations are large.  For this example, {\sc Maple}'s DifferentialAlgebra package was unsuccessful in finding IO equations within a few hours, but StructuralIdentifiability.jl was able to find them. This shows that our approach depends on symbolic tools to run in a tractable manner. 

%%%%%%%%%%%%%%%%%%%%%%%%%%%%%%%%%%%%%%%%%%%

\section{Examples}\label{sec:examples}
In this section, we will illustrate the approach 
\begin{itemize}
\item by a series of simple examples  that explain what can and cannot be achieved in principle and how the algorithm works in practice as well as
\item  by systems from modeling:
\begin{itemize} 
\item Lotka-Volterra  systems with and without inputs,
\item  a chemical reaction network system, whose reparametrization is non-scaling,  
\item a bilinear model with input, 
\item a rational Goodwin oscillator model,  
\item a rational blood coagulation and inhibition model with multiple outputs,  and
\item  a larger, Akt pathway model, with input and whose multiple outputs are combinations of state variables and parameters.
\end{itemize}
\end{itemize}
\subsection{Not possible to achieve global identifiability with any method}\label{sec:notglobal}
We will begin by showing that a  globally IO-identifiable reparametrization with the same order of the IO-equation does not always exist\footnote{We are grateful to Sebastian Falkensteiner and Rafael Sendra for pointing out a mistake in a previous version of this example and offering a correction.} (cf.~\cite[Theorem~3.5]{FPS2023}).
For this, consider the system
\begin{equation}\label{eq:impossible}
\begin{cases}
x' = \cfrac{\alpha_2}{2{\alpha_1}} (x^2 + 1),\\
y = \cfrac{2x}{{\alpha_2}(1 + x^2)}.
\end{cases}
\end{equation}
The corresponding IO-equation is \begin{equation}\label{IOeq:impossible}
{\alpha_1}^2(y')^2 + {\alpha_2}^2 y^2 - 1=0,
\end{equation}
and so the field of IO-identifiable functions is $\mathbb{Q}(\alpha_1^2, \alpha_2^2)$, and so neither ${\alpha_1}$ nor ${\alpha_2}$ are  globally {IO}-identifiable.
As in Theorem~\ref{thm:main}, we set $\beta_1 = \alpha_1^2$ and $\beta_2 = \alpha_2^2$.
If system~\eqref{eq:impossible} had had a reparametrization over $\mathbb{Q}({\beta_1, \beta_2})$ (and so had been IO-identifiable), 
then the curve $C$ (ellipse) defined by 
\[
{\beta_1} x^2+ {\beta_2} y^2=1
\] 
would  have had a rational parametrization over $\mathbb{Q}({\beta_1, \beta_2})$ but it does not. 
If it had had a parametrization over $\mathbb{Q}({\beta_1, \beta_2})$, it would have had a point 
\[
(x_0, y_0) = {(q_1(\beta_1, \beta_2), q_2(\beta_1, \beta_2)) \in \mathbb{Q}(\beta_1, \beta_2).}
\] 
We write both $q_1$ and $q_2$ as Laurent series in $\beta_1$ over the field $\mathbb{Q}(\beta_2)$.
Then $\beta_1 q_1^2$ is a Laurent series of odd valuation. 
Since $\beta_2 q_2^2$ and $1$ are of even valuation their 
valuations must be equal and the dominating terms must cancel.
Hence, \[q_2 = c_0(\beta_2) + \mathrm{O}(\beta_1),\] so the constant term of $1 - \beta_2q_2^2$ is $1 - \beta_2c_0(\beta_2)^2$.
This cannot be equal to zero because $1/\beta_2$ is not a square, so we arrive at a contradiction.

\subsection{Not possible to achieve global identifiability with this method}\label{sec:notglobthismethod}
 In this section, we give an example for which there is a reparametrization that gives global IO-identifiability, although our method only gives
a reparameterization with local IO-identifiability. 
Consider the  coupled by measurements exponential growth/decay system
\begin{equation}\label{eq:1}
\begin{cases}
x_1' = ax_1,\\
x_2' = bx_2,\\
y = x_1+ x_2,
\end{cases}
\end{equation}
and so $\bar x = (x_1,x_2)$, $\bar y = y$, and $\bar \alpha = (a,b)$. There is no $\bar u$.
The IO-equation is
\begin{equation}\label{eq:IOeqexab}
y'' - (a+b) y' + ab\cdot y = 0.
\end{equation}
Therefore, $\bar\beta = (a+b, a\cdot b)$ and the identifiable functions are $K := \mathbb{Q}(a+b, a\cdot b)$, and so $a$ and $b$ are algebraic of degree 2 over $K$, therefore, are only locally IO-identifiable.  Note that the coefficients of the IO-equation for
this example are the trace and determinant of the corresponding system’s 
matrix. For $i = 0, 1, 2$, we will compute $y^{(i)}$ as a function $R_i(x_1,x_2, a, b)$:
\begin{equation}
\begin{aligned}\label{eq:Liederexab}
y &= R_0(x_1,x_2, a, b) = x_1 + x_2,\\
y' &= R_1(x_1,x_2, a, b) = x_1'+x_2' = ax_1 + bx_2,\\
y'' &=R_2(x_1,x_2, a, b)= x_1''+x_2'' = a^2x_1 + b^2x_2,
\end{aligned}
\end{equation}
The Jacobian with respect to $\bar x$ is
\[
M = \begin{pmatrix}
1 & 1\\
a & b\\
a^2 & b^2
\end{pmatrix}
\]
Let $r = \operatorname{rank}  M = 2$ and ${D} = \det M = \begin{pmatrix}
1 & 1\\
a & b
\end{pmatrix} = b-a$ be a non-singular minor of $M$. 
Considering ${D}$ as a rational function in $\bar x$, we pick a (the only, in fact) non-zero coefficient ${D}_0(\bar\alpha)$ of its numerator as $b-a$. We now consider the following system of equations and inequations ($\ne$) in $\widetilde\alpha$ over $\mathbb{Q}(\bar\beta)$:
\[
\begin{cases}
a+b = \widetilde\alpha_1+\widetilde\alpha_2\\
a\cdot b = \widetilde\alpha_1\cdot\widetilde\alpha_2\\
\widetilde\alpha_1-\widetilde\alpha_2 \ne 0,
\end{cases}
\]
which has two solutions: $(\widetilde\alpha_1 = a, \widetilde\alpha_2 = b)$ and $(\widetilde\alpha_1 = b, \widetilde\alpha_2 = a)$,
neither of which make the locally identifiable model globally identifiable.

However, for example, the following reparametrization makes the model globally identifiable:
\[
\begin{cases}
w_1 =   x_1 + x_2,\\
w_2 =  ax_1 + bx_2,
\end{cases}
\]
resulting in this reparametrized  ODE system:
\[
\begin{cases}
w_1' = w_2,\\
w_2' = (a+b)\cdot w_2 - a\cdot b\cdot w_1,\\
y = w_1,
\end{cases}
\]
whose IO-equation is also~\eqref{eq:IOeqexab}.

\subsection{Choosing different solutions of~\eqref{eq:alphatilde}}\label{sec:choices}
Consider the  simple linear harmonic oscillator  system
\begin{equation}\label{eq:comparison}
\begin{cases}
x_1' = ax_2,\\
x_2' = bx_1,\\
y = x_1,
\end{cases}
\end{equation}
so $\bar x = (x_1,x_2)$, $\bar y = (y)$, $\bar\alpha = (a,b)$, and we have no $\bar u$.  Note that, for this system, to define a harmonic oscillator in the conventional sense, one requires $ab < 0$.
The IO-equation is
\begin{equation}\label{eq:comparison:IO}
y'' - ab\cdot y = 0.
\end{equation}
Therefore, $\bar\beta = (a b)$ and $a b$ is globally identifiable but neither $a$ nor $b$ is identifiable.  Note that the coefficient $ab$ of the IO-equation~\eqref{eq:comparison:IO} is the determinant of the matrix of~\eqref{eq:comparison}. 
Let us begin by computing Lie derivatives of $\bar y$. 
We have
\begin{equation}\label{eq:param:ex2}
\begin{aligned}
y &= R_0(x_1,x_2,a,b) =  x_1, \\
y' &= R_1(x_1,x_2,a,b) =x_1' = ax_2,\\
y''&= R_2(x_1,x_2,a,b) = x_1'' = ax_2' = abx_1.
\end{aligned}
\end{equation}
Following the the proof of Theorem~\ref{thm:main}, we have
\[
\beta_1 = ab = p_1(\bar\alpha) = p_1(a,b).
\] 
We then find the Jacobian of~\eqref{eq:param:ex2} with respect to $\bar x$:
\[
M = \begin{pmatrix}
1&0\\
0& a\\
ab & 0
\end{pmatrix}.
\]
Then $r := \operatorname{rank} M = 2 = \operatorname{trdeg}\mathbb{Q}(a,b)\langle\bar y\rangle/\mathbb{Q}(a,b)$. Let $D = \det\begin{pmatrix}
1&0\\
0& a
\end{pmatrix} = a$, a non-singular maximal minor of $M$. 
Considering $D$ as a rational function of $\bar x$, we pick a non-zero coefficient $D_0(\bar\alpha)$ of its numerator as $a$. 
We now consider the following system of  equations and inequations ($\ne$) over $\mathbb{Q}(\bar\beta)$:
\[
\begin{cases}
ab = \widetilde\alpha_1\cdot \widetilde\alpha_2,\\
\widetilde\alpha_1 \ne 0,
\end{cases}
\]
which has infinitely many solutions (over the algebraic closure of $\mathbb{Q}(ab)$),  including this one:  $\widetilde\alpha_1 = ab$, $\widetilde\alpha_2 =1$. Specializing~\eqref{eq:comparison}, we obtain the reparametrized system as:
\[
\begin{cases}
{w}_1' = {\beta_1}{w}_2,\\
{w}_2' = {w}_1,\\
y = {w}_1,
\end{cases}
\]
whose input-output equation is still~\eqref{eq:comparison:IO}, and the corresponding change of variables, calculated by equating the old and new Lie derivatives, is 
\[
w_1 =x_1,\quad w_2 =\frac{x_2}{b},
\]
which is a scaling reparametrization.
We could choose a different solution for $\widetilde{\alpha}_1, \widetilde{\alpha}_2$, for example, $\widetilde{\alpha}_1 = 1, \widetilde{\alpha}_2 = ab$ which would yield
\[
w_1' = w_2, \;\; w_2' = \beta_1w_1,\;\; y = w_1.
\]

\subsection{Lotka-Volterra examples}\label{sec:LV}
\subsubsection{Classical model}
Consider the system
\begin{equation}
\label{eq:LV}
\begin{cases}
x_1' = ax_1 -bx_1x_2,\\
x_2'= -cx_2 + dx_1x_2,\\
y = x_1
\end{cases}
\end{equation}
with two state variables $\bar{x} = (x_1,x_2)$, four parameters $\bar\alpha=(a,b,c,d)$, and one output $\bar y = y$. The input-output equation is
\begin{equation}
\label{eq:LVIO}
yy'' - y'^2 -dy^2y' + cyy' + ady^3 - acy^2 = 0.
\end{equation}
So, we have that the field of IO-identifiabile functions is
$\mathbb{Q}(d,c,ad,ac) = \mathbb{Q}(a,c,d)$.
The Lie derivatives of the $y$-variable are as follows:
\begin{equation}\label{eq:LieLV}
\begin{aligned}
y &= x_1, \\
y' &= -bx_1x_2 + ax_1,\\
y'' &= -bdx_1^2x_2 + b^2x_1x_2^2  + (bc-2ab)x_1x_2 + a^2x_1
\end{aligned}
\end{equation}
Following the proof of Theorem~\ref{thm:main}, we define
\[
\beta_1 = p_1(\bar\alpha)=a,\ \ \beta_2 = p_2(\bar\alpha)=c,\ \ \beta_3=p_3(\bar\alpha)=d.
\]
The Jacobian of~\eqref{eq:LieLV} w.r.t. $\bar x$ is
{\small \[
M = \begin{pmatrix}
1&0\\
a-bx_2&-bx_1\\
b^2x_2^2 - 2b(dx_1 + a - c/2)x_2 + a^2&-bx_1(dx_1 -2bx_2  + 2a - c)
\end{pmatrix}.
\]}
Then
\[
{D} = \det\begin{pmatrix}
1&0\\
a-bx_2&-bx_1
\end{pmatrix}=-bx_1
\]
is a maximal non-zero minor of $M$. Considering ${ D}$ as a rational function of $\bar x$, we pick a non-zero coefficient ${ D}_0(\bar\alpha)$ of its numerator as $-b$. We now arrive at the following system in $\widetilde\alpha$ over $\mathbb{Q}(\bar\beta)$:
\begin{equation}\label{eq:LVacd}
\begin{cases}
a=\widetilde\alpha_1,\\
c=\widetilde\alpha_3,\\
d=\widetilde\alpha_4,\\
-\widetilde\alpha_2\ne 0,
\end{cases}
\end{equation}
and we pick the following solution  (out of infinitely many solutions):
$\widetilde\alpha_1=a,
\widetilde\alpha_2=1,
\widetilde\alpha_3=c,
\widetilde\alpha_4=d
$, which results in the following reparametrized system of equations
\[
\begin{cases}
{ w}_1' = a{ w}_1 -{ w}_1{ w}_2,\\
{ w}_2'= -c{ w}_2 + d{ w}_1{ w}_2,\\
y = { w}_1
\end{cases}
\]
 The corresponding change of variables, obtained by equating old and new Lie derivatives, is
 \begin{equation}\label{eq:LVchange}
w_1=x_1,\quad w_2=bx_2,
\end{equation}
which is a scaling reparametrization in this case.
\subsubsection{Version with input}
Consider the following model from~\cite[eqs. (4)-(6)]{LVinput}:
\[
\begin{cases}
 x_1'=   ax_1 -bx_1x_2 + ux_1,\\
 x_2' = -cx_2 + dx_1x_2 + ux_2,\\
 y=x_1.
\end{cases}
\]
The IO-equation is
\begin{multline*}
yy''-y'^2-dy^2y'-yuy'+cyy'+duy^3+ady^3-y^2u'\\+y^2u^2+auy^2-cuy^2-acy^2=0.
\end{multline*}
Therefore, the {globally}  IO-identifiable parameters are $\mathbb{Q}(a,d,c)$. Following the steps from Section~\ref{sec:1}, we consider the Lie derivatives of $y$:
\begin{equation}\label{eq:LieLVu}
\begin{aligned}
y &= x_1, \\
y' &= x_1(-bx_2+a+u)
,\\
y'' &= x_1(x_2^2b^2-(dx_1+2a-c+3u)bx_2+(a+u)^2)
\end{aligned}
\end{equation}
The first $2\times 2$ principal minor of the Jacobian of~\eqref{eq:LieLVu} w.r.t. $\bar x$ is:
\[
\begin{pmatrix}
1&0\\
-bx_2+a+u & -bx_1
\end{pmatrix}.
\]
Then
\[
{D} = \det\begin{pmatrix}
1&0\\
-bx_2+a+u&-bx_1
\end{pmatrix}=-bx_1
\]
is a maximal non-zero minor of $M$. Considering ${ D}$ as a rational function of $\bar x$, we pick a non-zero coefficient ${ D}_0(\bar\alpha)$ of its numerator as $-b$. We again arrive at the system~\eqref{eq:LVacd} and we pick the following solution  :
$\widetilde\alpha_1=a,
\widetilde\alpha_2=1,
\widetilde\alpha_3=c,
\widetilde\alpha_4=d
$, which results in the following reparametrized system of equations
\[
\begin{cases}
w_1' = aw_1-w_1w_2+uw_1,\\
w_2' = -cw_2+dw_1w_2+uw_2.
\end{cases}
\]
 The corresponding change of variables, obtained by equating old and new Lie derivatives, is again~\eqref{eq:LVchange}.

\subsection{Chemical reaction network: non-scaling reparametrization}\label{sec:CRN}
Consider the following ODE model originating from a chemical reaction network, cf.~\cite[system~(3)]{DDP2023}
\[
\begin{cases}
X'=  k_2\cdot(A_{UX} + 2A_{XX} + A_{XU})  -k_1\cdot X\cdot (A_{UX} + A_{XU} + 2A_{UU}),\\
A_{UU}'=  k_2\cdot (A_{UX} + A_{XU}) -2k_1\cdot X\cdot A_{UU},\\
A_{UX}'= k_1\cdot X\cdot (A_{UU}-A_{UX}) + k_2 \cdot (A_{XX} - A_{UX}),\\
A_{XX}'=  k_1\cdot X\cdot (A_{UX} + A_{XU}) - 2\cdot k_2\cdot A_{XX},\\
A_{XU}'= k_1\cdot X\cdot(A_{UU}-A_{XU})+ k_2\cdot(A_{XX} - A_{XU}),\\
y = X.
\end{cases}
\]
We have $\bar x = (X, A_{UU}, A_{UX}, A_{XX}, A_{XU})$, $\bar\alpha = (k_1,k_2)$, $\bar y = (y)$, and there is no $\bar u$. A computation shows that
\begin{equation}\label{eq:crn_io}
y'y''' - (y'')^2 + 2k_1(y')^3=0
\end{equation}
is the IO-equation, and so $\bar\beta = (k_1)$ generates the field of IO-identifiable parameters.  So, $k_1$ is globally IO-identifiable and $k_2$ is IO-unidentifiable. 
Note that the order of the input-output equation is less than the dimension of the system, so we will first perform a preprocessing reduction as in Remark~\ref{rem:preprocessing}.

The Lie derivatives are as follows:
\begin{equation}\label{eq:oldLieCRN}
\begin{aligned}
y&=X\\
y'&=  k_2(A_{UX} + 2A_{XX} + A_{XU}) -k_1X(A_{UX}+2A_{UU}+A_{XU})\\
y''&=-y'\cdot(k_1X + 2k_1A_{UU} + k_1A_{UX} + k_1A_{XU} + k_2)\\
y'''&=y'\cdot\big((8Xk_1^2 + 4k_1(k_1A_{UX} + k_1A_{XU} + k_2))A_{UU} \\&\quad+(4k_1^2A_{UX} + 4k_1^2A_{XU} + 2k_1k_2)X\\
&\quad- 4k_1k_2A_{XX} + k_2^2  + k_1^2(A_{UX}^2 + 2A_{UX}A_{XU} + A_{XU}^2 + 4A_{UU}^2  + X^2)\big).\\
\end{aligned}
\end{equation}
We will now search for a three-dimensional system in only three variables $w_1, w_2, w_3$ having the same input-output equation. 
First we set up the desired Lie derivatives for this new system by replacing the original variables with arbitrary linear forms in $w_1, w_2, w_3$, say:
\[
  X = w_1,\; A_{UU} = w_2,\; A_{UX} = w_3, \; A_{XX} = A_{XU} = 0.
\]
We apply this substitution to~\eqref{eq:oldLieCRN}, equate the results before and after the substitution, and solve for $w_1, w_2, w_3$ (similarly to~\eqref{eq:xw}).
We get
\begin{equation}\label{eq:crn_trans}
  w_1 = X,\; w_2 = A_{UU} - A_{XX},\; w_3 = A_{UX} + 2 A_{XX} + A_{XU}.
\end{equation}
Then we can set up a linear system on $w_1', w_2', w_3'$ as in~\cite[Section~3]{Forsman} and find the reduced model:
\begin{equation}\label{eq:reducedCRN}
\begin{cases}
    w_1' = k_2w_3 - k_1 w_1 (w_3 + 2w_2),\\
    w_2' = -k_1 w_1 (w_3 + 2w_2) + k_2 w_3,\\
    w_3' = k_1 w_1 (w_3 + 2w_2) - k_2 w_3.
\end{cases}
\end{equation}
To this model, we can apply both parts of Theorem~\ref{thm:main}.
We still have the same IO-equation~\eqref{eq:crn_io}, so $\beta_1 = k_1$.
The first three Lie derivatives
\begin{align*}
  y &= w_1,\\
  y' &= k_2 w_3 - k_1 w_1 (w_3 + 2 w_2),\\
  y'' &= -y' (k_1(w_1 + 2 w_2 + w_3) k_1 + k_2)
\end{align*}
have nonsingular Jacobian, and one of the coefficients of its determinant is $k_1k_2^2$, so we set up a system
\[
\begin{cases}
    k_1 = \widetilde{\alpha}_1,\\
    \widetilde{\alpha}_1\widetilde{\alpha}_2 \neq 0.
\end{cases}
\]
We take a solution $\widetilde{\alpha}_1 = k_1$ and $\widetilde{\alpha}_2 = 1$  in $\overline{\mathbb{Q}(\bar\beta)}=\overline{\mathbb{Q}(k_1)}$ and, substituting this solution into~\eqref{eq:reducedCRN}, obtain
\[
\begin{cases}
    v_1' = v_3 - k_1v_1(v_3 + 2v_2),\\
    v_2' = -k_1v_1(v_3 + 2v_2) + v_3,\\
    v_3' = k_1 v_1 (v_3 + 2v_2) - v_3.
\end{cases}
\]
The corresponding state transformation from the last step of the proof of Theorem~\ref{thm:main} is
\begin{align*}
v_1 &= w_{1},\;\;\; v_3 = k_2 (w_1 + w_3) - w_1,\\
v_2 &= ((-w_1-w_3) k_2 + w_1 + 2 w_2 + w_3)/2 + \frac{k_2 - 1}{2k_1}.
\end{align*}
The overall state transformation between the new and original system can be obtained by composing this with~\eqref{eq:crn_trans}:
\begin{align*}
v_1 &= X,\\
v_2 &=X+2A_{UU}+A_{UX}+A_{XU}\\
&\quad-(X+A_{UX}+A_{XU}+2A_{XX})k_2+\frac{k_2-1}{2k_1},\\
v_3 &=\cfrac{X+A_{UX}+A_{XU}+2A_{XX}}{k_2}-X.
\end{align*}
\subsection{Bilinear model with input}
Consider the model~\cite[Example~1]{LLV}:
\[
\begin{cases}
x_1'= - p_1x_1 + p_2u,\\
x_2'=  - p_3x_2 + p_4u,\\
x_3'=- (p_1+p_3)x_3 + (p_4x_1+p_2x_2)u,\\
y=x_3.
\end{cases}
\]
Complete details of the computation are available here:~\cite{exaples-github}.
Computing the IO-equations, extracting the coefficients, and simplifying the IO-identifiable field generators, we obtain that the globally IO-identifiable functions are \[k := \mathbb{Q}(p_1p_3, p_2p_4, p_1+p_3).\] Following the steps from Section~\ref{sec:1} in our {\sc Maple} code for this example (e.g., a non-zero maximal minor of the Jacobian matrix of Lie derivatives is $p_2p_4(p_1-p_3)u^2$), we arrive at the following system of equations and inequations ($\ne$) in the unknowns $\widetilde{p_1},\widetilde{p_2},\widetilde{p_3},\widetilde{p_4}$: 
\[
\begin{cases}
p_1p_3 = \widetilde{p_1}\widetilde{p_3},\\
p_2p_4 = \widetilde{p_2}\widetilde{p_4},\\
p_1+p_3 = \widetilde{p_1}+\widetilde{p_3},\\
\widetilde{p_2}\widetilde{p_4}\left(\widetilde{p_1}-\widetilde{p_3}\right)\ne 0.

\end{cases}
\]
with solutions sought
in the field $\overline{k}$. We pick the tuple $(p_1, 1, p_3, p_2p_4)$ as a solution.
In this tuple, $1, p_2p_4 \in k$, and $p_1$ and $p_3$ are algebraic over $k$ since they satisfy a polynomial equation 
\begin{equation}\label{eq:polp1p3}
Z^2 - (p_1 + p_3)Z + p_1p_3 = 0
\end{equation}
with coefficients in $k$.
Thus, we arrive at the following ODE model, which is locally IO-identifiable:
\begin{equation}\label{eq:blrep}
\begin{cases}
w_1'= - p_1w_1 + u,\\
w_2'=  - p_3w_2 + p_2p_4u,\\
w_3'=- (p_1+p_3)w_3 + (p_2p_4w_1+w_2)u,\\
y=w_3,
\end{cases}
\end{equation}
in particular, all coefficients except for $p_1$ and $p_3$ are globally IO-identifiable (belong to $k$), and $p_1$ and $p_3$ are locally IO-identifiable because they are roots of the polynomial~\eqref{eq:polp1p3}, which defines the pair $(p_1,p_2)$ only up to a permutation. Equating the old and new Lie derivatives of the output variable, we obtain that~\eqref{eq:blrep} corresponds to the following change of variables:
\[
w_1 = \cfrac{x_1}{p_2},\quad w_2 = p_2x_2, \quad w_3 = x_3.
\]

\subsection{Goodwin oscillator  - rational model}\label{sec:GO}
Consider the following rational ODE model~\cite{goodwin_oscillatory_1965}:
\begin{equation}\label{eq:GOmodel}
\begin{cases}
 x_1' = -b x_1 + \cfrac{1}{c + x_4},\\
    x_2' = \alpha x_1 - \beta x_2,\\
    x_3' = \gamma x_2 - \delta x_3,\\
    x_4' =\cfrac{\sigma x_4 (\gamma x_2 - \delta x_3)}{ x_3},\\
    y = x_1.
\end{cases}
\end{equation}
We have $\bar x = (x_1,x_2, x_3,x_4)$, $\bar y = (y)$, $\bar\alpha = (b,c,\alpha,\beta,\gamma,\delta,\sigma)$, and there is no $\bar u$. We follow the steps from Section~\ref{sec:1} using {\sc Maple}, see~\cite{exaples-github}. The IO-equation is an order $4$ ODE in $y$ and is too big to display here. However, the field of IO-identifiable functions is 
\[
k := \mathbb{Q}(b,\;
c,\;
\sigma,\;
\beta\delta,\;
\beta+\delta
).
\] 
In particular, $b$, $c$, and $\sigma$ are globally IO-identifiable, $\beta$ and $\delta$ are locally but not globally IO-identifiable, and $\alpha$ and $\gamma$ are not locally IO-identifiable. The Lie derivatives are too big to display here too, but their Jacobian w.r.t. to $\bar x$ has a $4 \times 4$ minor equal
\begin{equation}\label{eq:GOM}
\cfrac{\alpha x_1x_4^2\gamma^2\sigma^2}{x_3^3(c + x_4)^6}.
\end{equation}
The numerator of  the fraction~\eqref{eq:GOM} has a non-zero coefficient $\alpha\gamma^2\sigma^2$. Therefore, we arrive at the following system of constraints:
\[
\begin{cases}
b = \widetilde{b},\ 
c = \widetilde{c},\ 
\sigma = \widetilde{\sigma},\ 
\beta\delta = \widetilde{\beta}\widetilde{\delta},\ 
\beta+\delta = \widetilde{\beta}+\widetilde{\delta},\\
\widetilde{\alpha}\widetilde{\gamma}^2\widetilde{\sigma}^2\ne 0.
\end{cases}
\]
Therefore, by the constructive proof of Theorem~\ref{thm:main} (Section~\ref{sec:1}), substituting any non-zero element of $k$ into $\alpha$ and $\gamma$ will turn~\eqref{eq:GOmodel} into a locally IO-identifiable model. For example, we can substitute (for simplicity) $\alpha=1$ and $\gamma = 1$:
\[
\begin{cases}
 w_1' = -b w_1 + \cfrac{1}{c + w_4},\\
    w_2' =  w_1 - \beta w_2,\\
    w_3' = w_2 - \delta w_3,\\
    w_4' =\cfrac{\sigma w_4 (w_2 - \delta w_3)}{ w_3},\\
    y = w_1,
\end{cases}
\]
and the corresponding change of variables can be found by equating the old and new Lie derivatives, and it is:
\[
w_1 = x_1,\quad
w_2 = \cfrac{x_2}{\alpha},\quad
w_3 = \cfrac{x_3}{\alpha\gamma},\quad
w_4 = x_4.
\]

\subsection{Rational model of blood coagulation and inhibition} Consider the following model from~\cite[eq.~(1)]{QXZXZX2004}:
\begin{equation}\label{eq:BCImodel}
\begin{cases}
   \IXa' = k_1\cdot \beta - h_1\cdot \IXa,\\
    \VIIIa' = k_2\cdot \IIa + \cfrac{k_3\cdot \APC\cdot \VIIIa}{b_1 + \VIIIa} - h_2\cdot \VIIIa,\\
    \Xa' = \cfrac{k_5\cdot  \IXa\cdot \VIIIa}{b2+\VIIIa} - h_3\cdot \Xa,\\
    \Va' = k_6\cdot  \IIa - \cfrac{k_7\cdot  \APC\cdot  \Va}{b3+\Va} - h_4\cdot \Va,\\
    \APC' = k_8\cdot  \IIa - h_5\cdot \APC,\\
    \IIa' = \cfrac{k_9\cdot  \Xa\cdot \Va}{b_4+\Va}-h_6\cdot \IIa,\\
    y_1 = \Xa,\\
    y_2 = \IIa,\\
    y_3 = \IXa,\\
    y_4 = \APC
\end{cases}
\end{equation}
The 4 IO-equations,  computed in {\sc Maple} ~\cite{exaples-github},  are too big to display here. The orders of the equations are $2$, $2$, $1$, and $1$, respectively. So, the sum of the orders is $6$. The field of IO-identifiable functions of the parameters is
\[
k := \mathbb{Q}\left(h_1,\ldots,
h_6,\;
k_5,\;
k_8,\;
k_9,\;
\tfrac{b_1}{k_3},\;
\tfrac{b_2}{k_3},\;
\tfrac{b_3}{k_7},\;
\tfrac{b_4}{k_7},\;
k_1\beta,\;
\tfrac{k_2}{k_3},\;
\tfrac{k_6}{k_7}\;
\right).
\]
The Jacobian matrix $M$ w.r.t. $\bar x$ of the Lie derivatives of $\bar y$ is too big to display as well. The first $6\times 6$ principal leading minor of $M$ is non-zero, and it has a non-zero coefficient w.r.t. $\bar x$ of the numerator equal to \[b_1^2b_2^2b_4^2\beta k_1k_5^2k_7k_9^2.\] We therefore, following Section~\ref{sec:1}, obtain the following system of constraints, after removing unnecessary exponents from the inequality product:
\begin{equation}\label{eq:constrBCI}
\begin{cases}
h_i = \widetilde{h}_i,\ 1\leqslant i\leqslant 6,\\
k_5 = \widetilde{k}_5,\
k_8 = \widetilde{k}_8,\
k_9 = \widetilde{k}_9,\\
\cfrac{b_1}{k_3} =\cfrac{\widetilde{b}_1}{\widetilde{k}_3},\
\cfrac{b_2}{k_3} =\cfrac{\widetilde{b}_2}{\widetilde{k}_3},\\
\cfrac{b_3}{k_7} =\cfrac{\widetilde{b}_3}{\widetilde{k}_7},\
\cfrac{b_4}{k_7} =\cfrac{\widetilde{b}_4}{\widetilde{k}_7},\\
k_1\beta = \widetilde{k}_1\widetilde{\beta},\\
\cfrac{k_2}{k_3} =\cfrac{\widetilde{k}_2}{\widetilde{k}_3},\
\cfrac{k_6}{k_7} =\cfrac{\widetilde{k}_6}{\widetilde{k}_7},\\
\widetilde{b}_1\widetilde{b}_2\widetilde{b}_4\widetilde{\beta}\widetilde{k}_1\widetilde{k}_5\widetilde{k}_7\widetilde{k}_9 \ne 0.
\end{cases}
\end{equation}
Our calculation in {\sc Maple} shows that the solution set
\begin{multline*}
\widetilde{b}_1 = \cfrac{b_1\widetilde{k}_3}{k_3},
\widetilde{b}_2 = \cfrac{b_2\widetilde{k}_3}{k_3},
\widetilde{b}_3 = \cfrac{b_3\widetilde{k}_7}{k_7},\
\widetilde{b}_4 = \cfrac{b_4\widetilde{k}_7}{k_7},
\widetilde{\beta} = \cfrac{k_1\beta}{\widetilde{k}_1},\\
\widetilde{h}_i = h_i,\ 1\leqslant i \leqslant 6,
\widetilde{k}_1 = \widetilde{k}_1,\ \widetilde{k}_2 = \cfrac{\widetilde{k}_3k_2}{k_3}, \\\widetilde{k}_3 = \widetilde{k}_3, \widetilde{k}_5 = k_5, \widetilde{k}_6 = \cfrac{\widetilde{k}_7k_6}{k_7}, 
\widetilde{k}_7 = \widetilde{k}_7, \widetilde{k}_8 = k_8, \widetilde{k}_9 = k_9
\end{multline*}
of~\eqref{eq:constrBCI} can be parametrized by $\left(\widetilde{k}_1,\widetilde{k}_3,\widetilde{k}_7\right)$, with none of $\widetilde{k}_1$,  $\widetilde{k}_3$, or $\widetilde{k}_7$ being zero. As one of the reparametrization choices, we let $\widetilde{k}_1=\widetilde{k}_3=\widetilde{k}_7=1$, obtaining the following locally identifiable reparametrization of~\eqref{eq:BCImodel}:
\begin{equation}\label{eq:BCImodel-rep}
\begin{cases}
w_1' = K_1-h_1w_1,\\
w_2' = \cfrac{\left(w_5-B_1h_2+K_2w_6\right)w_2+K_2B_1w_6-h_2w_2^2}{B_1+w_2},\\
w_3' = \cfrac{k_5w_1w_2-h_3w_2w_3-B_2h_3w_3}{B_2+w_2},\\
w_4' = \cfrac{\left(K_6w_6-B_3h_4-w_5\right)w_4+K_6B_3w_6-h_4w_4^2}{B_3+w_4},\\
w_5' = k_8w_6-h_5w_5,\\
w_6' = \cfrac{k_9w_3w_4-h_6\left(B_4+w_4\right)w_6}{B_4+w_4},\\
   y_1 = w_1,\\
    y_2 = w_3,\\
    y_3 = w_5,\\
    y_4 = w_6,
\end{cases}
\end{equation}
where 
\[K_1 = k_1\beta,\ B_1 = \tfrac{b_1}{k_3},\ K_2 = \tfrac{k_2}{k_3},\ B_2 = \tfrac{b_2}{k_3},\ K_6 = \tfrac{k_6}{k_7},\ B_3 = \tfrac{b_3}{k_7},\ B_4 = \tfrac{b_4}{k_7}\]
are all in the field $k$.
By setting the Lie derivatives of the outputs from~\eqref{eq:BCImodel} equal to those of~\eqref{eq:BCImodel-rep}, we arrive at the following change of variables to achieve the reparametrization~\eqref{eq:BCImodel-rep}:
\begin{align*}
    w_1 &= \IXa, &w_2 &= \cfrac{\VIIIa}{k_3}, &w_3 &= \Xa,\\
     w_4 &= \cfrac{\Va}{k_7}, &w_5 &= \APC, &w_6 &= \IIa.
\end{align*}

\subsection{ Larger example: Akt pathway model with input}\label{sec:Akt}
 As a larger example to test and compare the efficiency of different steps of our algorithm, consider the Akt pathway model~\cite{Fujita}. This is a non-linear model with 16~parameters, 9~state variables, 1~input, and 3~outputs, as in \cite[Example~11]{dong2023differential}:
\begin{equation}\label{eq:Akt}
\begin{cases}
X_1' = {X_1}_t  u+X_9  {r_1}_{ k_2}-X_1  {X_1}_t-X_9  {r_1}_{ k_1}\\
X_2' = -X_2  X_4  {r_2}_{ k_1}-X_2  {r_4}_{ k_1}+X_3  {r_2}_{ k_2}+X_3  {r_3}_{ k_1}+X_9  {r_9}_{ k_1}\\
X_3' = X_2  X_4  {r_2}_{ k_1}-X_3  {r_2}_{ k_2}-X_3  {r_3}_{ k_1}\\
X_4' = -X_2  X_4  {r_2}_{ k_1}+X_3  {r_2}_{ k_2}+X_5  {r_7}_{ k_1}\\
X_5' = -X_5  X_6  {r_5}_{ k_1}+X_3  {r_3}_{ k_1}-X_5  {r_7}_{ k_1}+X_7  {r_5}_{ k_2}+X_7  {r_6}_{ k_1}\\
X_6' = -X_5  X_6  {r_5}_{ k_1}+X_7  {r_5}_{ k_2}+X_8  {r_8}_{ k_1}\\
X_7' = X_5  X_6  {r_5}_{ k_1}-X_7  {r_5}_{ k_2}-X_7  {r_6}_{ k_1}\\
X_8' = X_7  {r_6}_{k_1}-X_8  {r_8}_{k_1}\\
X_9' = X_9  {r_1}_{k_1}-X_9  {r_1}_{k_2}-X_9  {r_9}_{ k_1}\\
y_1 = a_1  (X_2+X_3)\\
y_2 = a_2  (X_5+X_7)\\
y_3 = a_3  X_8
\end{cases}
\end{equation}
The computation of IO equations did not finish in {\sc Maple}. Instead, we used the Julia package \href{https://github.com/SciML/StructuralIdentifiability.jl}{StruncturalIdentifiability.jl} \cite{dong2023differential}, which was able to find IO equations (see the \href{https://github.com/SciML/StructuralIdentifiability.jl/blob/master/examples/Fujita.jl}{Fujita.jl} example), but they are too big to display here but are written explicitly in the corresponding {\sc Maple} worksheet~\cite{exaples-github}, which we also used in the rest of this example.
After computing IO equations  and simplifying, we find that the IO-identifiable functions of  the parameters are
\begin{multline}
\label{eq:Akt:ident}
{r_2}_{k_2},\
{r_3}_{k_1},\
{r_4}_{k_1},\
{r_5}_{k_2},\
{r_6}_{k_1},\
{r_7}_{k_1},\
{r_8}_{k_1},\\
\frac{a_2}{{r_5}_{k_1}},\
\frac{a_3}{{r_5}_{k_1}},\
\frac{{r_2}_{k_1}}{{r_5}_{k_1}},\
\frac{{r_5}_{k_1}}{a_1},\
{r_1}_{k_1}-{r_1}_{k_2}-{r_9}_{k_1}.
\end{multline}
The combined order of the IO equations is $8$, while there are $9$ state variables. Moreover, the variable $X_1$ does not appear in the Lie derivatives. Therefore, we can only guarantee a locally IO-identifiable reparametrization of system~\eqref{eq:Akt} with the first equation excluded. However, we will still attempt to reparametrize the entire system.

A computation shows that, after computing Lie derivatives and picking a non-zero $8\times 8$ minor of the Jacobian (as in the algorithm), we obtain
\[
D = {r_3}_{k_1}{r_9}_{k_1}{r_6}_{k_1}^3{r_5}_{k_1}^2{{r_4}_{k_1}}{r_2}_{k_1}a_2a_1^3a_3^4X_2X_5X_6
\]
and then 
\begin{equation}
\label{eq:Atk:forinequation}
D_0(\bar\alpha) = {r_3}_{k_1}{r_9}_{k_1}{r_6}_{k_1}^3{r_5}_{k_1}^2{{r_4}_{k_1}}{r_2}_{k_1}a_2a_1^3a_3^4.
\end{equation}
Based on~\eqref{eq:Akt:ident} and~\eqref{eq:Atk:forinequation}, we obtain the following polynomial system:
\[
\begin{cases}
\widetilde{r_2}_{k_2} = {r_2}_{k_2},\ \ 
\widetilde{r_3}_{k_1}={r_3}_{k_1},\ \
\widetilde{r_4}_{k_1}={r_4}_{k_1},\ \
\widetilde{r_5}_{k_2}={r_5}_{k_2},\\
\widetilde{r_6}_{k_1}={r_6}_{k_1},\ \
\widetilde{r_7}_{k_1}={r_7}_{k_1},\ \
\widetilde{r_8}_{k_1}={r_8}_{k_1},\\
\frac{\widetilde a_2}{\widetilde{r_5}_{k_1}}=\frac{a_2}{{r_5}_{k_1}},\ \
\frac{\widetilde a_3}{\widetilde{r_5}_{k_1}}=\frac{\widetilde a_3}{\widetilde{r_5}_{k_1}},\ \ 
\frac{\widetilde{r_2}_{k_1}}{\widetilde{r_5}_{k_1}}=\frac{{r_2}_{k_1}}{{r_5}_{k_1}},\ \
\frac{\widetilde{r_5}_{k_1}}{\widetilde a_1}=\frac{{r_5}_{k_1}}{a_1},\\
{\widetilde{r_1}}_{k_1}-{\widetilde{r_1}}_{k_2}-{\widetilde{r_9}}_{k_1}={r_1}_{k_1}-{r_1}_{k_2}-{r_9}_{k_1},\\
{\widetilde{r_3}}_{k_1}{\widetilde{r_9}}_{k_1}{\widetilde{r_6}}_{k_1}^3{\widetilde{r_5}}_{k_1}^2{{\widetilde{r_4}}_{k_1}}{\widetilde{r_2}}_{k_1}\widetilde{a_2}\widetilde{a_1}^3\widetilde{a_3}^4 \ne 0.
\end{cases}
\]
We pick the following solution (can be found by hand or in {\sc Maple}):
\begin{equation}\label{eq:solsAkt}
\begin{gathered}
\widetilde{{X_1}_t} = 1,\ \ 
\widetilde{a_1} = \frac{a_1}{{r_5}_{k_1}},\ \ 
\widetilde{a_2} = \frac{a_2}{{r_5}_{k_1}}\\
\widetilde{a_3} = \frac{a_3}{{r_5}_{k_1}},\ \ 
\widetilde{r_1}_{k_1} = {r_1}_{k_1}-{r_1}_{k_2}-{r_9}_{k_1}, \ \ 
\widetilde{r_1}_{k_2} = -1\\
\widetilde{r_2}_{k_1} = \frac{{r_2}_{k_1}}{{r_5}_{k_1}},\ \ 
\widetilde{r_2}_{k_2} = {r_2}_{k_2}, \ \ 
\widetilde{r_3}_{k_1} = {r_3}_{k_1}\\
\widetilde{r_4}_{k_1} = {r_4}_{k_1},\ \ 
\widetilde{r_5}_{k_1} = 1, \ \ 
\widetilde{r_5}_{k_2} = {r_5}_{k_2}\\
\widetilde{r_6}_{k_1} = {r_6}_{k_1}, \ \
\widetilde{r_7}_{k_1} = {r_7}_{k_1},\ \ 
\widetilde{r_8}_{k_1} = {r_8}_{k_1}, \ \ 
\widetilde{r_9}_{k_1} = 1
\end{gathered}
\end{equation}
Replacing each parameter in~\eqref{eq:Akt} with the value for the corresponding solution with tilde and $X$'s with $w$'s in~\eqref{eq:solsAkt}, we obtain the following reparametrized ODE system:
\begin{equation}\label{eq:Akt:new}
\begin{cases}
w_1' = (-{r_1}_{k_1}+{r_1}_{k_2}+{r_9}_{k_1}-1) w_9 +u -w_1 \\
w_2' = \left(-  \frac{{r_2}_{k_1}}{{r_5}_{k_1}}w_4-{r_4}_{k_1}\right) w_2 +w_3  ({r_2}_{k_2}+{r_3}_{k_1})+w_9 \\
w_3' = (-{r_2}_{k_2}-{r_3}_{k_1}) w_3 + \frac{{r_2}_{k_1}}{{r_5}_{k_1}}w_2 w_4 \\
w_4' =   {r_7}_{k_1}w_5+  {r_2}_{k_2}w_3-  \frac{{r_2}_{k_1}}{{r_5}_{k_1}}w_2  w_4\\
w_5' = (-w_6 -{r_7}_{k_1}) w_5 +({r_5}_{k_2}+{r_6}_{k_1}) w_7 +  {r_3}_{k_1}w_3\\
w_6' =   {r_5}_{k_2}w_7+  {r_8}_{k_1}w_8-w_5  w_6 \\
w_7' = (-{r_5}_{k_2}-{r_6}_{k_1}) w_7 +w_5  w_6 \\
w_8' =  {r_6}_{k_1} w_7-  {r_8}_{k_1}w_8\\
w_9' =   ({r_1}_{k_1}-{r_1}_{k_2}-{r_9}_{k_1})w_9\\
y_1 = \frac{a_1}{{r_5}_{k_1}}  (w_2+w_3)\\
y_2 = \frac{a_2}{{r_5}_{k_1}}  (w_5+w_7)\\
y_3 = \frac{a_3}{{r_5}_{k_1}}  w_8
\end{cases}
\end{equation}
Moreover, a calculation shows that all of the parameters in the reparametrized system~\eqref{eq:Akt:new} with all $9$ state variables are globally identifiable.
The corresponding change of variables is:
\begin{gather*}
w_2 = {r_5}_{k_1}X_2\ \ \
w_3 = {r_5}_{k_1}X_3\ \ \
w_4 = {r_5}_{k_1}X_4\ \ \ 
w_5 = {r_5}_{k_1}X_5\\
w_6 = {r_5}_{k_1}X_6\ \ \
w_7 = {r_5}_{k_1}X_7\ \ \
w_8 = {r_5}_{k_1}X_8\ \ \
w_9 = {r_5}_{k_1}{r_9}_{k_1}X_9
\end{gather*}
where we could also manually set $w_1 = X_1$ as our choice because neither $X_1$ nor $w_1$ appear in the Lie derivatives.

\section{Conclusions}

We described an algorithm for input-output locally identifiable reparametrizations of ODE models, proved its correctness, and gave a range of illustrating examples. More future work is still needed to develop a fully algorithmic approach that ensures identifiability of the initial conditions and works with non-smooth or differential algebraic inputs as well as inputs with delays, which occur in practice.

\section*{Acknowledgments}We are grateful to the CCiS at CUNY Queens College for the computational resources and to Julio Banga, Sebastian Falkensteiner, Gemma Massonis, Nikki Meshkat, Rafael Sendra and Alejandro Villaverde for useful discussions and the referees for careful reading and helpful comments.

This work was partially supported by the NSF grants CCF-2212460, CCF-1563942, CCF-1564132, DMS-1760448, DMS-1760212, DMS-1760413, DMS-1853650, and DMS-1853482 and CUNY grant PSC-CUNY \#65605-00 53.

\bibliographystyle{abbrvnat}
\bibliography{bib}

\end{document}